\documentclass[preprint,showpacs,amsmath,amssymb,prb]{revtex4}
\usepackage{graphicx}

\begin{document}

\title{Broadband sensitive pump-probe setup for ultrafast optical switching of photonic nanostructures and semiconductors}

\author{Tijmen G. Euser$^{1,2,3}$}

\author{Philip J. Harding$^{1,2}$}

\author{Willem L. Vos$^{1,2}$}

\email{w.vos@amolf.nl} \homepage{www.photonicbandgaps.com}

\affiliation{$^1$FOM Institute for Atomic and Molecular Physics (AMOLF),
Kruislaan 407, 1098 SJ Amsterdam, The Netherlands}

\affiliation{$^2$ Complex Photonic Systems, MESA$^+$ Institute for Nanotechnology, University of Twente, The Netherlands}

\affiliation{$^3$Max Planck Institute for the Science of Light, Gu\"nther-Scharowsky-Str. 1, Bau 24
91058 Erlangen, Germany }

\pacs{42.70.Qs, 42.65.Pc, 42.79.-e}

\begin{abstract}
We describe an ultrafast time resolved pump-probe spectroscopy setup aimed at studying the switching of nanophotonic structures. Both fs pump and probe pulses can be independently tuned over broad frequency range between 3850 and 21050 cm$^{-1}$. A broad pump scan range allows a large optical penetration depth, while a broad probe scan range is crucial to study strongly photonic crystals. A new data acquisition method allows for sensitive pump-probe measurements, and corrects for fluctuations in probe intensity and pump stray light. We observe a tenfold improvement of the precision of the setup compared to laser fluctuations, allowing a measurement accuracy of better than $\Delta$R= 0.07$\%$ in a 1 s measurement time. Demonstrations of the improved technique are presented for a bulk Si wafer, a 3D Si inverse opal photonic bandgap crystal, and z-scan measurements of the two-photon absorption coefficient of Si, GaAs, and the three-photon absorption coefficient of GaP in the infrared wavelength range.
\end{abstract}

\maketitle

\section{Introduction}
\label{introduction}

Optical pump-probe experiments \cite{Demtroder02,Siegman86,Diels96} are a powerful tool to study the ultrafast optical response of a wide range of effects in, for example, semiconductor physics,\cite{Muecke0} high harmonic generation,\cite{Esry06} and optics of biomembranes.\cite{Roke03} Recently, pump-probe techniques have also been extended to study ultrafast switching of photonic nanostructures such as photonic crystals \cite{Leonard02,Mazurenko03,Becker05,Euser07} and photonic cavities.\cite{Fushman07,Preble07,Harding07} In these studies, one literally attempts to catch light with light. Therefore such switching processes are susceptible to perturbing effects such as absorption or (induced) inhomogeneity,\cite{Euser05} and sensitive experimental methods are required.

In pump-probe experiments, high pump pulse intensities are often required to observe small changes in probe reflection or transmission. This requirement has led to the development of regenerative amplifiers, in which femtosecond (fs) pulses from Ti:Sapphire lasers are amplified to pulse energies of up to several mJ. The amplification process comes at a price of a strongly reduced repetition rate, typically from the MHz to the kHz range. The amplification step is often followed by conversion to different wavelengths using optical parametric amplifiers (OPA). Amplification processes typically increase pulse-to-pulse intensity variations of the laser.

There are two important issues that limit the speed and accuracy of pump-probe experiments.  First, since experiments intrinsically depend in the magnitude of the irradiance, they are sensitive to pulse-to-pulse variations of the laser. The result is that long integration times are required to sufficiently reduce the fluctuation-induced error in probe reflectivity measurement. Secondly, scattered light from the intense pump pulses contributes to the background signal of the probe detector. In particular for strongly photonic samples, light is necessarily strongly scattered, therefore the background level can be larger than the reflectivity changes of the samples under study. To circumvent both potential issues, we have developed a versatile measurement scheme that allows for compensation for pulse-to-pulse variations in the output of our laser, as well as a subtraction of the pump background from the probe signal. Our technique strongly reduces the acquisition times required in pump-probe experiments, allowing for much more detailed scans than previously possible. While this paper focuses on the application to switching of semiconductors and nanophotonic structures through optical excitation of free carries, our results are relevant to any pump-probe experiment with regeneratively amplified laser pulses.

\section{pump-probe setup}
\label{setup}

\subsection{Optical setup}
\label{setup:optical}
Time-resolved optical measurements on photonic crystals were performed with a dedicated two-color pump-probe setup. Our laser system provides high power pulses at two independently tunable frequencies, allowing us to adjust the pump frequency to optimize the optical penetration depth\cite{Euser05} and the probe frequency to scan across broad photonic gaps. The setup is based on a regeneratively amplified Titanium Sapphire laser that emits short 120 fs pulses at $\lambda$= 800 nm with a pulse energy of 1 mJ at a repetition rate of 1 kHz (Spectra Physics Hurricane). This laser drives two optical parametric amplifiers (OPA, Topas 800-fs) shown schematically in Fig.~\ref{fig:setup}, that serve as pump and probe. The output frequencies of the OPAs are computer controlled, and can be continuously tuned between 3850 and 21050 cm$^{-1}$. The excitation of carriers at pump frequencies near the two-photon absorption edge of semiconductors requires a high pump irradiance in the range of 10 to 300 GWcm$^{-2}$, depending on the material and the pump frequency chosen.\cite{Euser05} Since both OPAs have a conversion efficiency that exceeds 30$\%$, a pulse energy $E_{pulse}$ of at least 20 $\mu$J is available over the entire frequency range. The output of our OPAs consists of pulses with Gaussian pulse duration $\tau_p$= 140$\pm$10 fs (measured at $\lambda$= 1300 nm). The spectral shape of the output spectrum was measured to be Gaussian with a frequency independent linewidth ${\Delta\nu/\nu}$= 1.44$\pm$0.05$\%$. We deduce the time-bandwidth product to be $\tau_p\Delta\nu$= 0.47$\pm$0.05, in good agreement with the Fourier limit for Gaussian pulses ($\tau_p\Delta\nu$= 0.44).\cite{Siegman86} To a good approximation, the temporal profile of the pulses has a Gaussian intensity envelope:

\begin{figure}[!ht]
\begin{center}
\includegraphics[width=0.8\linewidth]{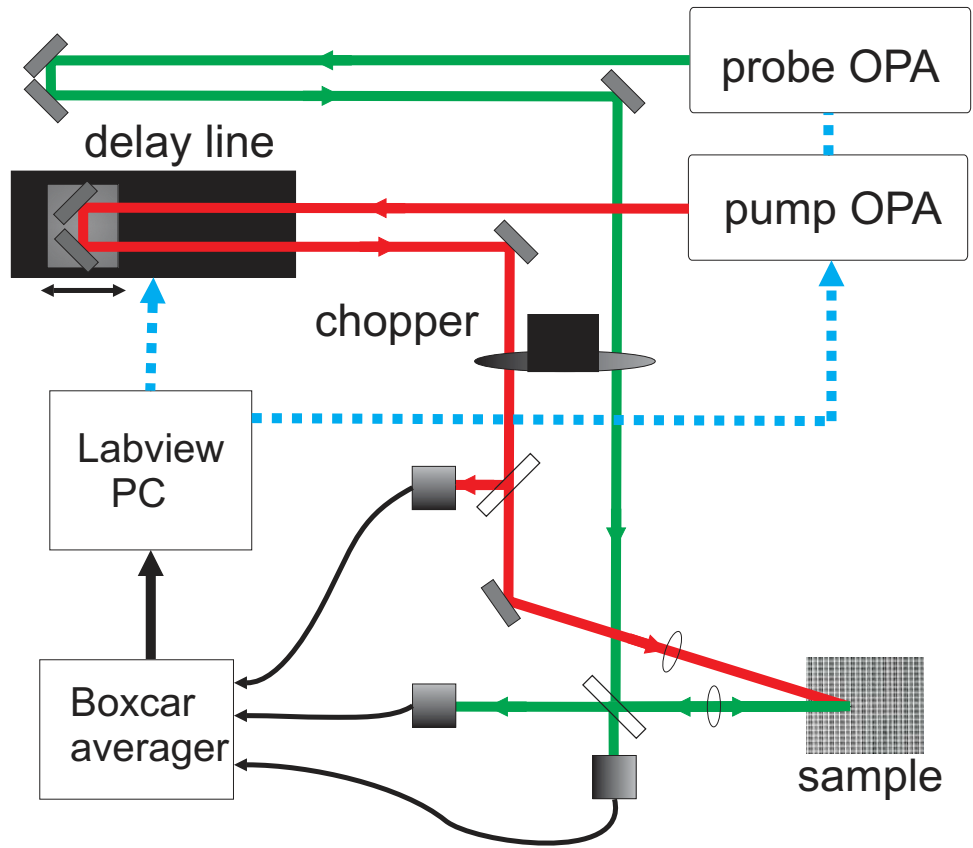}\\
\caption{\label{fig:setup} (Color online) Schematic drawing of the setup. Pulses (Gaussian pulse duration $\tau_p$= 120 fs, $\lambda$= 800 nm, E= 1 mJ) from a regeneratively amplified Ti:Saph laser (not shown) drive two OPAs. The output frequency of both OPAs is computer controlled, and tunable from 3850 to 21050 cm$^{-1}$.  The pump pulse passes through an optical delay line with minimum time step of 10 fs. Both pump and probe beam pass through a chopper wheel that is synchronized to the laser output (see Fig.~\ref{fig:chopper}). Both pump and probe beam are focused on the same spot on the sample. Two InGaAs photodiodes are used to monitor the output variation of the OPAs as well as the reflected signal. The reflectivity from the sample is measured by a third InGaAs photodiode. Three boxcar averagers are used to hold the short output pulses of each detector for 1 ms, allowing simultaneous acquisition of separate pulse events of all three detector channels.}
\end{center}
\end{figure}

\begin{equation}
\label{eq:tau}
P(t)=P_{max}e^{-2\big(\frac{t}{\tau_p}\big)^2},
\end{equation}

\noindent where $P_{max}=( E_{pulse}/\tau_p)(\sqrt{2/\pi})$ is the peak power. Experimentally, we obtain the $P_{max}$ inside the sample by subtracting the pump reflectivity at the sample interface:

\begin{equation}
\label{eq:I0} P_{max}=P_{ext}(1-R),
\end{equation}

\noindent where $P_{ext}$ is the external pump power, and R is the measured reflectivity of the pump beam at the sample interface at $\lambda_{pump}$.  Both pump and probe beams were focused onto the sample at a small numerical aperture NA= 0.02. The pump intensity profile was confirmed to be Gaussian with a radius $w_{pump}$= 113$\pm$5 $\mu$m. We can therefore describe the spatial irradiance distribution in the focus as Gaussian:

\begin{equation}
I(x,y)= I_0 e^{-2\frac{x^2+y^2}{w_{pump}^2}},
\end{equation}

\noindent where $I_0=(P_{max}/w_{pump}^2)(2/\pi)$ is the peak irradiance in the center of the focus. Even with a large pump focus of $w_{pump}$= 113$\pm$5 $\mu$m, the maximum peak irradiance $I_{max}$ that can be obtained in our setup still exceeds 1 TWcm$^{-2}$. This large excess irradiance indicates that it is feasible to switch an even larger sample, or to use less powerful lasers on small samples, which is important to facilitate possible future applications of ultrafast switching.

The probe beam was focused to a Gaussian spot of typical radius $w_{probe}$= 20$\pm$5$\mu m$, depending on the diffraction limited size of the spot given by $\lambda_{probe}$. Since the probe focus is much smaller than the pump focus, we obtain excellent lateral homogeneity of the nonlinear excitation throughout the probe focus. This turns out to be crucial to permit successful physical interpretation of complex photonic structures. In all experiments, we explicitly ensured that only the central flat part of the pump focus is probed by testing with a Si wafer.

In Figure~\ref{fig:setup}, the delay between pump- and probe pulse was set by a 40 cm long optical delay line with a time resolution of $\Delta$t= 10 fs. Since the delay time is also computer controlled, we can scan the reflectivity spectrum as a function of frequency at a chosen time delay after the pump pulse.

\subsection{Data acquisition method}

\begin{figure}[!ht]
\begin{center}
\includegraphics[width=0.5\linewidth]{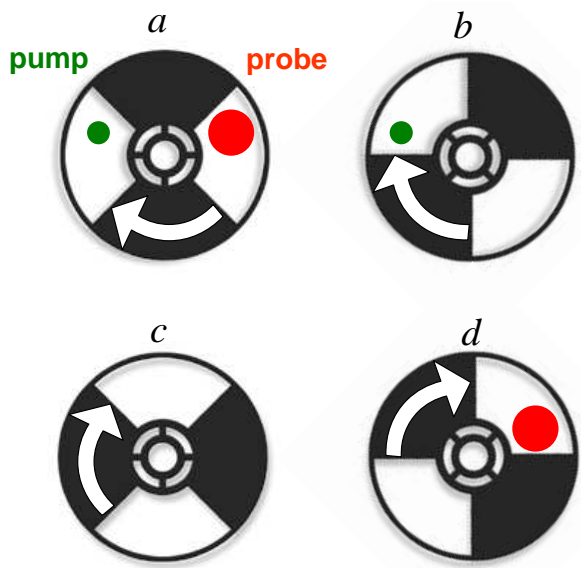}\\
\caption{\label{fig:chopper} (Color online) Schematic drawing illustrating the alignment of the pump- en probe beams (red and green circles) onto the chopper blade. The rotation of the chopper wheel is synchronized to the laser output. One full revolution of the chopper blade takes 8 ms, such that that for each pulse event, pump and probe beams are blocked or unblocked in a different permutation. In one sequence of four consecutive laser pulses, both (a) excited reflectivity, (d) linear reflectivity,(b) pump background, and (c) detection background are collected.}
\end{center}
\end{figure}

In our setup, the OPAs show typical relative irradiance variation between 2$\%$ RMS variation near $\lambda$= 1300 nm and of 7$\%$ in the worst case near the degeneracy point near $\lambda$= 1600 nm. If the signals are not corrected for, such variations in pump irradiance in a one-photon process would fundamentally limit the relative accuracy of the probe signal. To improve the signal-to-noise to better than the laser stability, it is important to probe individual pulse events so that pulse selection can be performed. Therefore, the irradiance of each pump and probe pulse is monitored by two InGaAs photodiodes and the reflectivity signal is measured by a third InGaAs photodiode, shown as black squares in Fig.~\ref{fig:setup}. Three boxcar averagers are used to hold the short output pulses of each detector for 1 ms, allowing simultaneous acquisition of separate pulse events of all three detector channels by a data acquisition card. Both pump and probe beam pass through a chopper whose frequency is synchronized to the repetition rate of the laser ( $\Omega_{\rm{rep}}$= 1 kHz). The alignment of the two beams on the chopper blade is such that each millisecond, pump and probe beam are blocked or unblocked in a different permutation, shown in Fig.~\ref{fig:chopper}. In this flexible measurement scheme, detector signals for each pulse event are collected, allowing various data processing routines such as automatic background subtraction and the selection of pulses within a certain pump energy range after the experiment.

For a measurement on a reflecting sample, the linear (unpumped) reflectance is given by $R^{\rm{up}} = J^{\rm{up}} - J^{\rm{up}}_{\rm{bg}}$, where $J^{\rm{up}}$ is the detector signal when the chopper is in position (d), while $J^{\rm{up}}_{\rm{bg}}$ is the probe background signal measured at chopper position (c). To compensate for probe pulse fluctuations, $R^{\rm{up}}$ is then ratioed by the background-corrected probe monitor signals $M^{\rm{up}}$, measured when the chopper is at position (d) and (c). As the background measurements are taken in between the reflectivity measurements, temporal fluctuations in the background signal originating from pump and from the surroundings are eliminated. In a similar manner, the non-linear (pumped) reflectance is equal to $R^{\rm{p}} = J^{\rm{p}} - J^{\rm{p}}_{\rm{bg}}$, where $J^{\rm{p}}$ and $J^{\rm{p}}_{\rm{bg}}$ are the signals measured on R at chopper positions (a) and (b), respectively. This signal is also ratioed to the corresponding probe monitor signals. This process obviously requires the three detectors to store all four signals during a time ($4/\Omega_{\rm{rep}}$). When this happens, the differential reflectivity $\Delta R/R$ corrected for background and fluctuations is thus determined by

\begin{equation}
\frac{\Delta R}{R} \equiv \frac{R^{\rm{p}}/M^{\rm{p}} -
R^{\rm{up}}/M^{\rm{up}}}{R^{\rm{up}}/M^{\rm{up}}}.
\end{equation}

The signal $J$ is offered to the DAC card by the boxcar measuring the sample reflectance. Neglecting electronic amplification factors, $J$ is equal to the magnitude of the time- and space integrated Poynting vector ${\bf{S}}$,

\begin{eqnarray}
J = \pi r^2\int_{-t_{\rm{int}}/2}^{t_{\rm{int}}/2}|{\bf{S}}| dt &=&
\int_{-t_{\rm{int}}/2}^{t_{\rm{int}}/2} \sqrt{\frac{\epsilon_0}{\mu_0}}\digamma(t)^2 dt
\nonumber \\ &\approx& \pi r^2 \sqrt{\frac{\epsilon_0}{\mu_0}} \frac{\tilde{\digamma}_0^2}{2}
\int_{-\infty}^{\infty}\left(\exp(-t^2/\tau_P^2)\right)^2dt \nonumber \\ &=&\pi \sqrt{\pi} r^2
\sqrt{\frac{\epsilon_0}{\mu_0}}\frac{\tau_P \tilde{\digamma}^2_0}{4},
\label{eq:SVEA}
\end{eqnarray}

Here, the beam is collimated to a radius $r$ and $t_{\rm{int}}$ is the integration time of the boxcar. $\epsilon_0$ and $\mu_0$ denote the permittivity and permeability of free space, respectively. The electric field $\digamma(t)$ reflected by a sample onto the detector can be separated in a Gaussian envelope $\tilde{\digamma}(t)$ of temporal width $\tau_P$ (see Eq.~\ref{eq:tau}) and amplitude $\tilde{\digamma}_0$, multiplied by a sinusoidal component with a carrier frequency $\omega_0$ in rad/s.\footnote{This Slowly Varying Envelope Approximation (SVEA, see, e.g. Ref.~\onlinecite{Diels96}) can be applied to pulses where $\tau_P >> 1/\omega_0$, and where $\omega_0$ does not change over $\tau_P$, i.e., for bandwidth limited pulses.} The squared oscillating term can then be integrated separately and yields $1/2$, and the time integration can be taken to infinity because $t_{\rm{int}} >> \tau_P$. Since the integration time of the boxcar ($t_{\rm{int}} \sim$ 150 ns) is much longer than any probe pulse duration, the dynamics of the sample is essentially integrated over.

\begin{figure}[!ht]
\begin{center}
\includegraphics[width=0.7\linewidth]{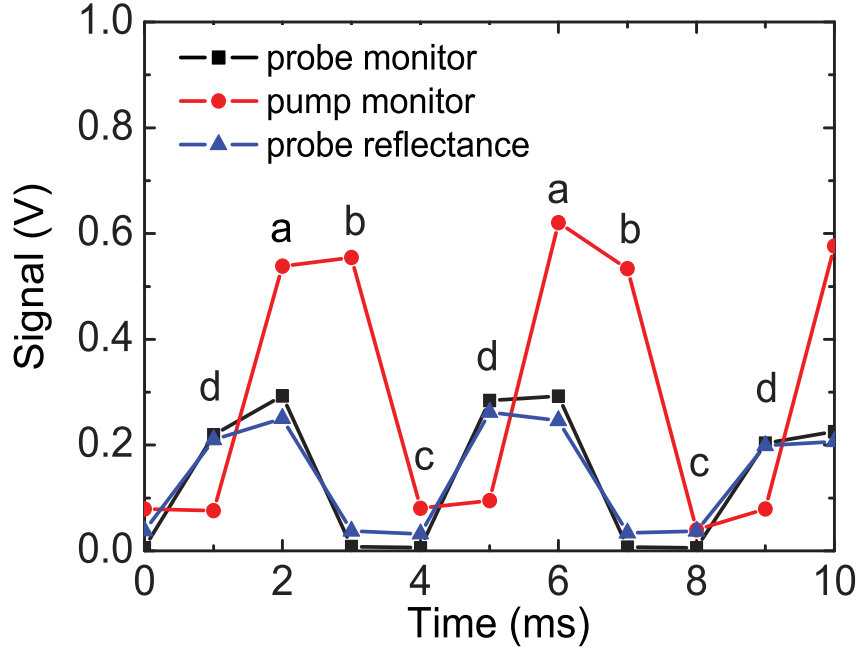}\\
\caption{\label{fig:signals} (Color online) Time traces of the boxcar output signals for probe monitor, pump monitor, and probe reflectance. The sample was a GaAs/AlAs distributed Bragg reflector, the experimental conditions were the same as in Ref.~\onlinecite{Euser07b}. The pump irradiance was $\approx$100 GWcm$^{-2}$ and the probe frequency was $\lambda$= 1490 nm. The switched reflectivity is roughly 10$\%$ lower than the unswitched reflectivity. Each datapoint in the plot corresponds
to a single pulse event. The letters a, b, c, and d correspond to the chopper position during each event (see Fig.~\ref{fig:chopper}).}
\end{center}
\end{figure}

Figure~\ref{fig:signals} shows a typical time trace for probe monitor, pump monitor and reflected probe signal collected by the data acquisition card. Note that between 1 ms (pump off) and 2 ms (pump on) the probe reflectance signal goes up, suggesting an increase in reflectivity, while between 5 and 6 ms, the probe reflectance signal decreases. These artifacts are caused by the pulse-to-pulse variations in the laser output, and are easily eliminated in our method by referencing to the probe monitor signal. An additional advantage of our scheme is that excited and linear reflectivity signals can be simultaneously monitored on an oscilloscope, which greatly facilitates the alignment procedure.

\section{Experimental results}
\label{results}

In this section we will demonstrate how our technique yields precise nonlinear reflection and transmission measurements, both on intricate photonic crystal samples as well as on bulk semiconductors.

\subsection{Statistical analysis of measured data}

\begin{figure}[!ht]
\begin{center}
\includegraphics[width=0.7\linewidth]{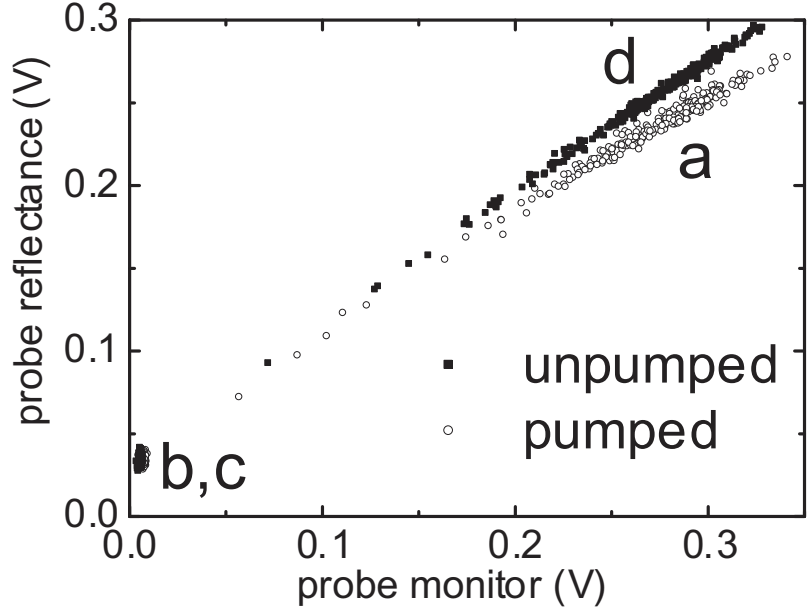}\\
\caption{\label{fig:boxcar} Reflectance signal versus probe monitor data for 1000 single pulse events of the data set shown in Fig.~\ref{fig:signals}, displayed as a scatter plot. The 250 unpumped reflectivity datapoints (d) constitute a line, indicating that variations in monitor and reflectance signal are strongly correlated. The slope of the line is proportional to the reflectivity of the sample. The pumped datapoints (a) form a line with a reduced slope, due to the reflectivity decrease of about $\delta$R/R= 10$\%$ in the switched sample. Both background data sets (b) and (c) tend to the origin of the plot as it should in absence of offsets. Note that the small offset in the signals is automatically removed in the data processing routine.}
\end{center}
\end{figure}

In this section we describe the statistical analysis of the data collected in our experiments. At each time delay and for each wavelength setting for the probe OPA, all detector signals from 4x250 pulse events were collected and stored. The probe reflectance signal for 4x250 pulse events of a pump-probe experiment on a GaAs/AlAs multilayer structure is plotted versus probe monitor signal in Figure~\ref{fig:boxcar}. Experimental details for this structure can be found in Ref.~\onlinecite{Euser07b}. Both signals show a variation as a result of the pulse-to-pulse variations of the laser. The datapoints constitute two separate lines whose slopes correspond to the unpumped and pumped reflectance of the sample. To exemplify the noise reduction of our method, we have chosen a data set during which the alignment of the pump laser was not optimized and pulse-to-pulse variations of the probe signal were larger than normal, amounting to a large relative standard deviation
$\sigma_{SD,probe}$= 13$\%$.

The corresponding standard error in the mean detector signal is $\delta$R/R=$\sigma_{SD,probe}$/$\sqrt{N}$= 13$\%$/$\sqrt{250}$= 0.8$\%$, which is relatively large compared to the effects that we wish to study. We therefore use an automated data processing routine to process the probe reflectance and probe monitor data to increase the signal-to-noise ratio. From the entire data set, the averages of the background levels (b) and (c) were determined, and subtracted from the pumped (a) and unpumped (d) reflectance data respectively (see Fig.~\ref{fig:chopper}). The resulting background-subtracted reflectance signal was divided by the corresponding monitor signal to compensate for intensity variations in the output of the laser. Through this procedure, the RMS variation in the unpumped reflectivity that was found from the data in Fig.~\ref{fig:boxcar} was strongly reduced to $\sigma_{SD,probe}$= 1.1$\%$. We attribute the remaining noise to uncorrelated electronic noise in the detection system. The resulting standard error in the probe reflectivity is thus tenfold improved to $\delta$R/R= 0.07$\%$, even if the laser is not running optimally. Our scheme allows a sensitive measurement of the reflectivity and of small reflectivity changes, while maintaining an acceptable measurement time of about 1 second per frequency-delay setting.

Pulse to pulse variations in the pump energy are a more subtle issue, since such fluctuations will often propagate in a nonlinear, and sometimes unpredictable, way in the reflectivity change $\Delta$R/R of the sample. The open circles in Fig.~\ref{fig:boxcar} correspond to the switched reflectance data. The slope of the line that is formed by these data points is reduced by about 10$\%$ compared to the unswitched data (closed squares). The corresponding reflectivity decrease is equal to $\Delta$R/R= 10$\%$. We also observe that the line is about twice as broad as the line corresponding to the unpumped data. We attribute the broadening to pulse to pulse variations of the pump beam.

In the example in Fig.~\ref{fig:boxcar} the standard deviation of the pump pulse energy $\sigma_{SD, pump}$= 12$\%$, from which we deduce a standard error in the pump irradiance $\delta I_0/I_0$= $\frac{\sigma_{SD, pump}}{\sqrt{N}}$= 0.8$\%$.  The error in the reflectivity change $\Delta$R/R due to the pulse-to-pulse irradiance fluctuations is equal to $\delta$($\Delta$R/R)= 2($\delta I_0/I_0)(\Delta$R/R), where the factor 2 is due to the quadratic dependence of $\Delta$R/R on $I_0$ for a two-photon process. Using the reflectivity change in the data in Fig.~\ref{fig:boxcar} ($\Delta$R/R= 10$\%$) we obtain the pump contribution to the standard error in $\Delta$R/R to be $\delta$($\Delta$R/R)= 0.16$\%$. The error in reflectivity changes $\Delta$R/R also contains a contribution of the fluctuations in the probe pulse that were discussed before. The error due to the probe variation is equal to $\sqrt{2}(\delta$R/R)= 0.1$\%$, since the two independent errors in the pumped- and unpumped datasets are added. We calculate the total error by adding the contributions of both probe and pump variations. We obtain a standard error:

\begin{equation}
\label{eq:daq} \delta{\Big (}\frac{\Delta R}{R}{\Big  )}
=\sqrt{2}\frac{\sigma_{SD,probe}}{\sqrt{N}}+\frac{\Delta
R}{R}\frac{\sigma_{SD,pump}}{\sqrt{N}}= 0.26\%,
\end{equation}

\noindent which is sufficiently sensitive for our switching experiments.

In some applications, an even higher sensitivity is required. Fortunately, pump-monitor detector signals for each individual pulse event are stored. It is thus possible to reduce the pump term in Eq.~\ref{eq:daq} by selecting pump pulses within a certain narrow energy range after the experiment, at the expense of longer integration times. Alternatively, in experiments where the relation between pump intensity and sample response is linear, the pump-monitor signal can be used to correct the measured signal. In our switching experiments on photonic nanostructures, however, such a correction cannot be made since the sample response is typically nonlinear with pump intensity. Therefore a pulse selection procedure was applied in z-scan measurements (see Section~\ref{results:zscan}), where pump stability is essential for the correct interpretation of the experimental data. Our strongly improved sensitivity has recently allowed us to identify two femtosecond contributions to the spectral properties of a switched Si woodpile photonic bandgap crystal: the optical Kerr effect, and nondegenerate two-photon absorption.\cite{Harding08}

\subsection{Ultrafast switching of bulk Si}
\label{results:siwafer}

\begin{figure}[!ht]
\begin{center}
\includegraphics[width=0.7\linewidth]{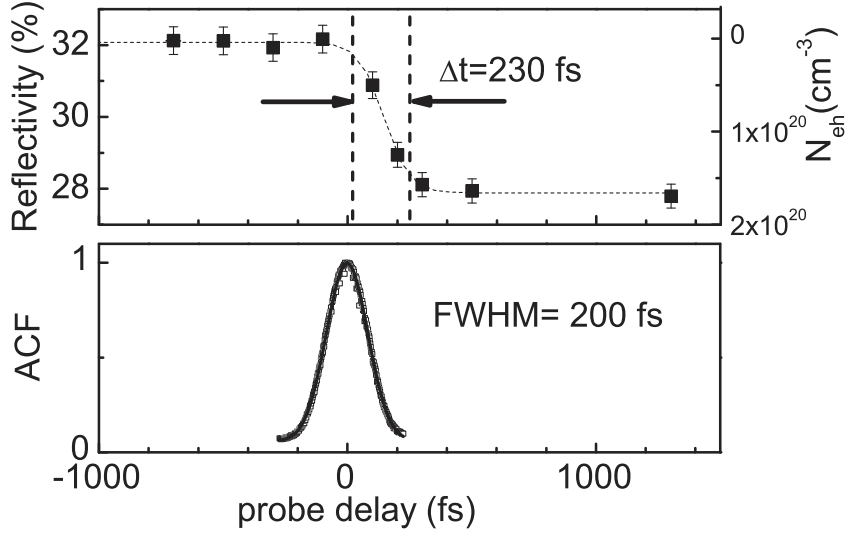}\\
\caption{\label{fig:siwafer} Time resolved reflectivity measurement on bulk Si, pumped at $\lambda_{pump}$= 800, pulse energy E$_{pump}$= 2.0$\pm$0.1 $\mu$J, Gaussian pulse duration $\tau_{pump}$= 120$\pm$10 fs, w$_{pump}$= 70$\pm$10 $\mu$m and peak irradiance 115$\pm$40 GWcm$^{-2}$ (upper panel). The reflectivity of a probe with $\lambda_{probe}$= 1300 nm, w$_{probe}$= 20$\pm$10 $\mu$m and Gaussian probe pulse duration $\tau_{probe}$= 120$\pm$10 fs decreases from 32$\%$ to 28$\%$, corresponding to a calculated carrier density N$_{eh}$= 1.6$\times$10$^{20}$ cm$^{-3}$ at the surface of the sample, using a Drude response (see right-hand scale). The time difference between 10$\%$ and 90$\%$ of the total change is 230$\pm$40 fs, as indicated by the vertical dashed lines. The lower panel shows the irradiance autocorrelation function (ACF) of the pump pulses. The full width half maximum (FWHM) of the ACF of 200 fs corresponds to a Gaussian pulse duration of $\tau_p$= 120$\pm$10 fs.}
\end{center}
\end{figure}

An example of free carrier-induced change of refractive index in bulk silicon is given in Fig.~\ref{fig:siwafer} (upper panel). In this experiment, a powerful ultrashort pump pulse with wavelength $\lambda_{pump}$= 800 nm was focused to a spot with radius w$_{pump}$=70$\pm$10 $\mu$m, resulting in a peak irradiance at the sample interface of I$_{0}$= 115$\pm$40 GWcm$^{-2}$. The reflectivity of a weaker probe pulse ($\lambda_{probe}$= 1300 nm) with a smaller spot radius of w$_{probe}$= 20$\pm$5 $\mu$m was measured in the center of the pumped spot at different time delays with respect to the pump pulse. The scan in Fig.~\ref{fig:siwafer} (upper panel) shows that the reflectivity of the sample changes from $32\%$ to $28\%$.
The $10\%-90\%$ rise time is 230 fs, clearly an ultrafast change in refractive index $n'$. From Fresnel's formula we find the refractive index change to be more than $10\%$. The lower panel shows the intensity autocorrelation function (ACF) of the pump pulses. The full width half maximum (FWHM) is 200 fs, we therefore conclude that the free carriers have been generated almost instantaneously.

\begin{figure}[!ht]
\begin{center}
\includegraphics[width=0.7\linewidth]{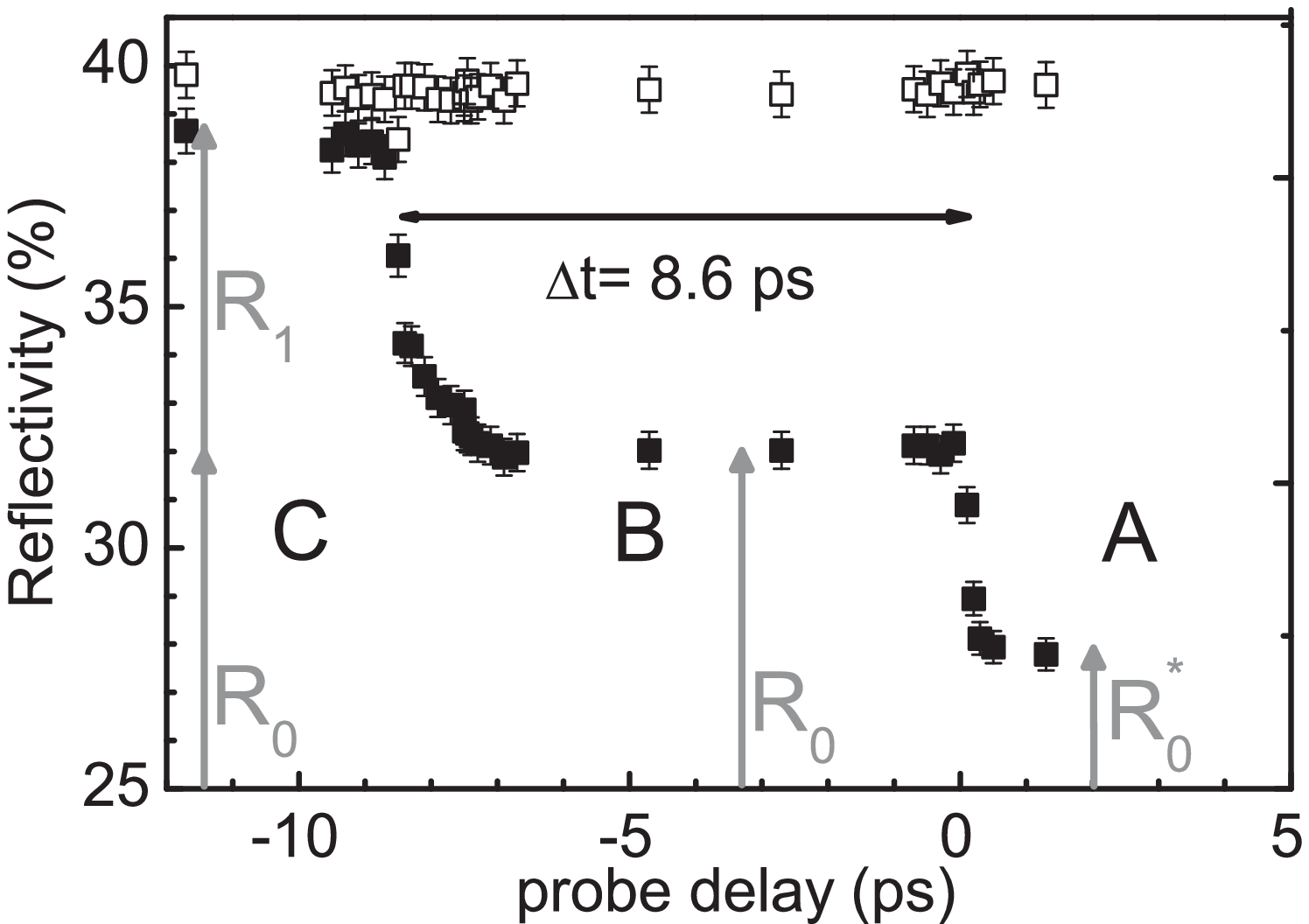}\\
\caption{\label{fig:siwaferlong} Time resolved reflectivity of a switched double side polished Si wafer. Unswitched reflectivity (open squares) and switched reflectivity (closed squares) are plotted over an extended range of probe delays compared to Fig.~\ref{fig:siwafer}. Surprisingly, at a negative probe delay of 8.6 ps, a large step in the reflectivity from 38$\%$ to 32$\%$ appears. At zero probe delay the reflectivity decreases further from 32$\%$ to 28$\%$. The time difference between the first and second step  in reflectivity (indicated by dotted lines) is 8.6$\pm$0.5 ps, which corresponds well to twice the optical thickness of the wafer (8.3$\pm$0.1 ps). We identify three different probe delay regimes A, B, and C, that are explained in the schematic plot in
Fig.~\ref{fig:siwaferschematic}.}
\end{center}
\end{figure}

Fig.~\ref{fig:siwaferlong} shows reflectivity from an extended probe delay range of -12 to +5 ps. Quite remarkably, at a negative probe delay of 8.6 ps, we observe an additional large step in the reflectivity from 38$\%$ to 32$\%$. We can identify three distinct probe delay regimes A, B, and C, which are separated by two large steps in the reflectivity. The time difference between the first and second step is 8.6$\pm$0.5 ps, this value corresponds well
to twice the optical thickness of the wafer $2Ln_{Si}/c$= 8.3$\pm$0.1 ps, where $n_{Si}$= 3.5 is the refractive index of Si at $\lambda$= 1300 nm,\cite{Palik85} and $L$= 356$\pm$5 $\mu$m is the measured thickness of the wafer.

\begin{figure}[!ht]
\begin{center}
\includegraphics[width=0.8\linewidth]{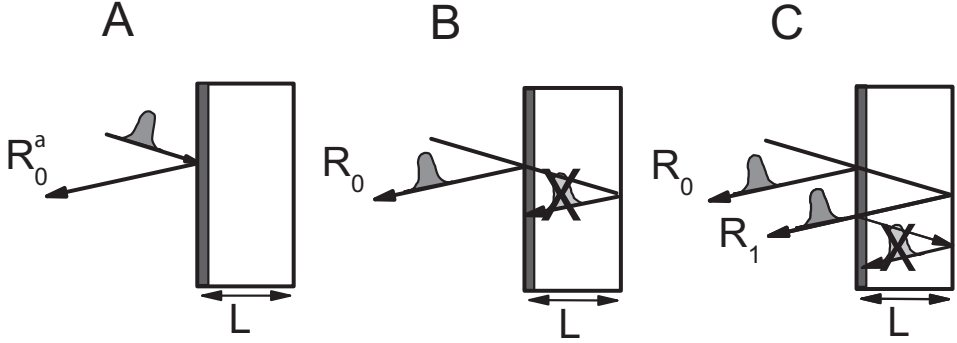}\\
\caption{\label{fig:siwaferschematic} Snap shots of the reflected irradiance of a bulk Si wafer in the experiment in Fig.~\ref{fig:siwaferlong} at the arrival time of the pump. We consider three different probe delay positions, corresponding to the regions indicated by A, B, and C in Fig.~\ref{fig:siwaferlong}. The intense pump pulse generates an inhomogeneous carrier plasma near the front face of the wafer indicated by the dark gray layer. This absorbing layer acts as an ultrafast shutter that blocks any internally reflected pulse $R_m$ that arrives at the front face after the pump pulse. At probe delay A, the pump arrives before the probe, and the switched reflectivity of the front face of the wafer $R_0^*$ is probed. At probe delay B, the pump pulse arrives in between reflection $R_1$ and $R_0$, thus blocking pulses $R_1$ and $R_2$. This reflection corresponds to the front face refection of the unswitched wafer $R_0$. At probe delay C, the pump pulse arrives in between reflection $R_2$ and $R_1$, thus only blocking $R_2$. The total reflection is equal to $R_0$+$R_1$.}
\end{center}
\end{figure}

To interpret the observed unusual time dependence, we show in Fig.~\ref{fig:siwaferschematic} snapshots of the reflected irradiance, taken at the moment that the pump pulse switches the front face of the wafer. The reflectivity of the wafer consists of multiply reflected pulses from front and back surface of the wafer, which are indicated by $R_0$, $R_1$, and $R_2$. The magnitude of each successive reflection is given by $R_m= (1-R_0)^2R_0^{2m-1}$. We neglect any reflections beyond $R_2$ in our analysis. It is important to note that each subsequent reflection $R_m$ is delayed with respect to $R_0$ by an even multiple (2m) of the optical thickness $\Delta$t$_m$= 2m$Ln_{Si}/c$.

The pump conditions in the experiment in Fig.~\ref{fig:siwaferlong} result in an inhomogeneous, dense carrier plasma near the front face of the wafer.\cite{Euser05} Free-carrier absorption and diffraction from the dense plasma results in a strongly attenuated transmission.  The plasma thus acts as an ultrafast shutter that blocks internally reflected pulses $R_m$ that arrive at the front face after the switching. At probe delay A, the pump arrives before reflection $R_0$, and the measured signal corresponds to the reflection of the switched wafer, indicated by $R_0^*$. At delay setting B, the pump pulse arrives in between reflection $R_1$ and $R_0$, thus blocking $R_1$ and $R_2$. This reflection corresponds to the front face refection of the unswitched wafer $R_0$. In Fig.~\ref{fig:siwaferlong} we observe that the reflection in this time range is indeed comparable to the Fresnel reflection from a single air-Si interface (R= 31$\%$). At probe delay C, the pump pulse arrives in between reflection $R_2$ and $R_1$, and only $R_2$ is blocked. The total reflection is thus equal to $R_0$+$R_1$. We note that the reflectivity changes in a switched double side polished wafer are very large, particularly when the back face reflection $R_1$ is blocked. Double side polished Si wafers are therefore ideal test samples to find- and optimize the spatial- and temporal overlap of our pump and probe pulses.

\subsection{Ultrafast switching of 3D Si inverse opal photonic bandgap crystals}
\label{results:inverseopal}

\begin{figure}[!ht]
\begin{center}
\includegraphics[width=0.7\linewidth]{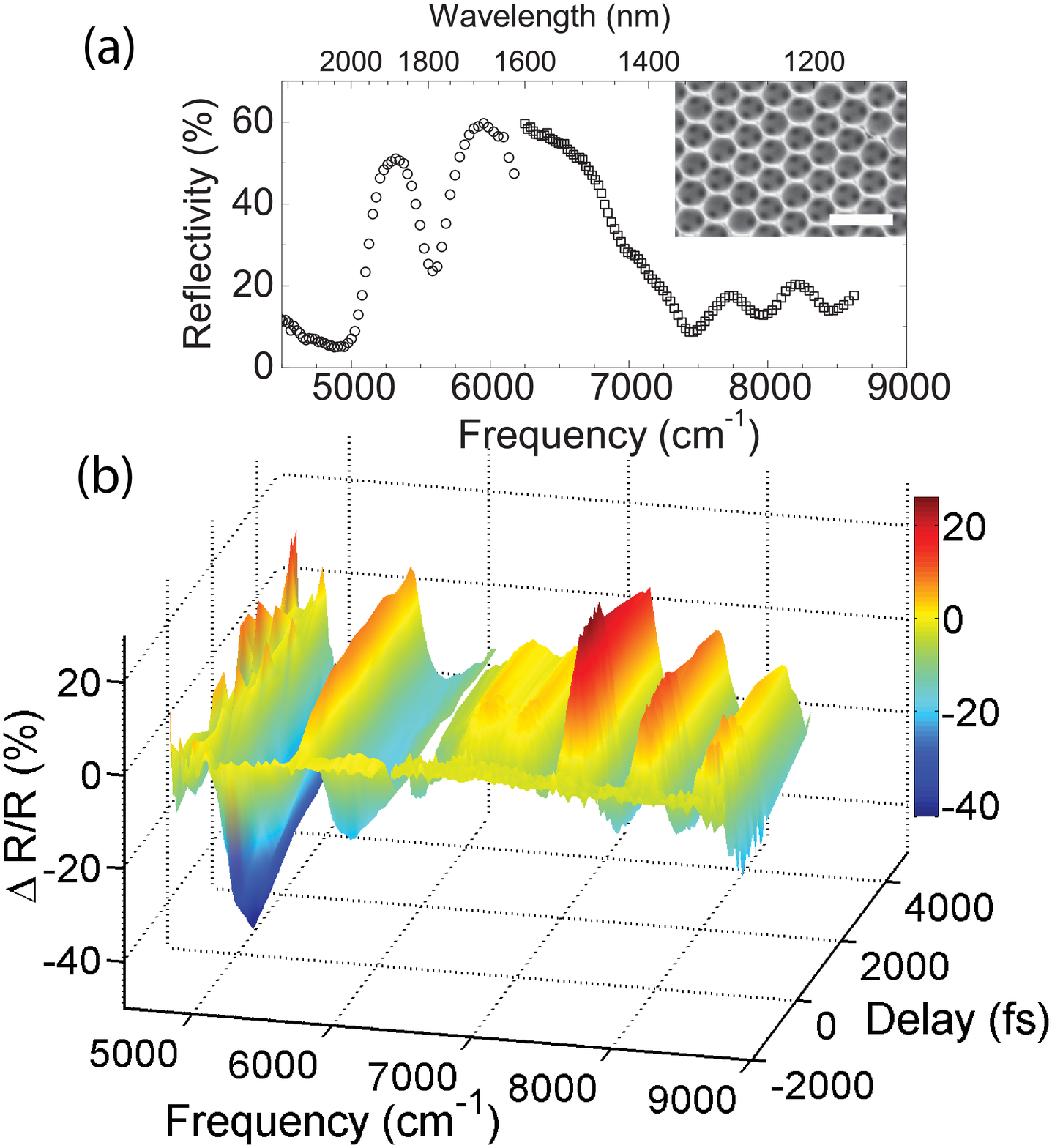}\\
\caption{(Color online) (a) Broadband linear reflectivity spectrum in the \emph{(111)} direction of a Si
inverse opal, measured by combining the signal and idler range of our optical amplifier.  Inset: High resolution SEM image of the Si inverse opal. The scale bar is 2 $\mu$m. Image courtesy of Jeroen Kalkman. (b) Differential reflectivity as a function of both probe frequency $\omega_{probe}$ and probe delay. The pump frequency and peak irradiance were $\lambda_{pump}$= 1550 nm and $I_0$= 4$\pm$1 GWcm$^{-2}$ on the red part, and $\lambda_{pump}$= 2000 nm and $I_0$= 25$\pm$3 GWcm$^{-2}$ on the blue part of the spectrum. The probe delay was varied in small steps of $\Delta$t= 50 fs on the blue edge and in steps of $\Delta$t= 500 fs at the red edge. The probe wavelength was tuned from 1600 to 2100 nm in $\Delta\lambda$= 10 nm steps in the low frequency range, and from 1160 to 1600 nm in 5 nm steps in the high-frequency range.\label{fig:f1133dplot}}
\end{center}
\end{figure}

As a second example we demonstrate ultrafast switching experiments that were carried out on the Si inverse opal photonic crystal shown in Fig.~\ref{fig:f1133dplot}(a) (inset). The broadband reflectivity data, shown in Fig.~\ref{fig:f1133dplot}(a) covers the complete range of second order stop bands in our crystal where a 3D photonic band gap has been predicted.\cite{Vlasov01} A two-photon process was used to homogeneously excite carriers in the photonic crystal. The pump frequency was chosen in relation to the probe frequency range, to allow polarization based separation of pump and probe light. The time and frequency resolved differential reflectivity of the crystal $\Delta$R/R($\tau,\omega_{probe}$) at ultrafast time scales is represented as a three-dimensional surface plot in Fig.~\ref{fig:f1133dplot}(b). The plot contains over 1500 datapoints, each obtained from 500 or 1000 single pulse measurements and represents nearly an octave on probe frequency. Such a large scanning range is typically required for strongly photonic crystals, since their photonic gaps have large bandwidths. The data collection was performed in as little as 40 minutes, much shorter than in conventional setups without automated scanning, -referencing and -background subtraction.\cite{Leonard02,Mazurenko03,Becker05} The data show clear dispersive shapes in the differential reflectivity, caused by a shift of the peaks towards higher frequency. We observe a large frequency shift of up to $\Delta\omega/\omega$= 1.5$\%$ of all spectral features including the peak that corresponds to the photonic band gap. We deduce a corresponding large refractive index change of $\Delta n_{Si}/n_{si}$= 2.0$\%$, where $n_{Si}$ is the refractive index of the silicon backbone of the crystal. Our broad probe scanning range allowed us to observe that both the low and high frequency edge of the stop bands have shifted. This indicates the absence of separate dielectric and air bands in the range of second order Bragg diffraction in inverse opals, which is consistent with predictions based on quasistatic band structure calculations. A detailed description of the ultrafast switching of Si inverse opal photonic bandgap crystals is presented in Ref.~\onlinecite{Euser07}.

\subsection{Z-scan measurements on GaP at IR wavelengths.}
\label{results:zscan}

As a final demonstration of our technique, we have obtained open aperture z-scan data for the semiconductors GaP, Si, and GaAs. A detailed description of these experiments is given in Appendix~\ref{appendix:zscan}. In this section we present measurements of the three-photon absorption coefficient of GaP, a highly suitable material for photonic bandgap crystals because of its high refractive index and low absorption in the visible wavelength range. The pump wavelength was chosen in the range of three-photon absorption: $\frac{1}{3}E_{gap} < \hbar\omega < \frac{1}{2}E_{gap}$. We therefore neglect both linear and two-photon absorption in our analysis ($\alpha$=$\beta$= 0). The resulting equation for the nonlinear transmission of the sample, normalized to the linear transmission $(1-R)^2$ is:

\begin{equation}
T(z)= \frac{1}{\sqrt{1+2I_0(z)^2\gamma L}}
\label{eq:threephoton},
\end{equation}

\noindent where $\gamma$ is the three-photon absorption coefficient.

\begin{figure}[!ht]
\begin{center}
(A).\includegraphics[width=0.4\linewidth]{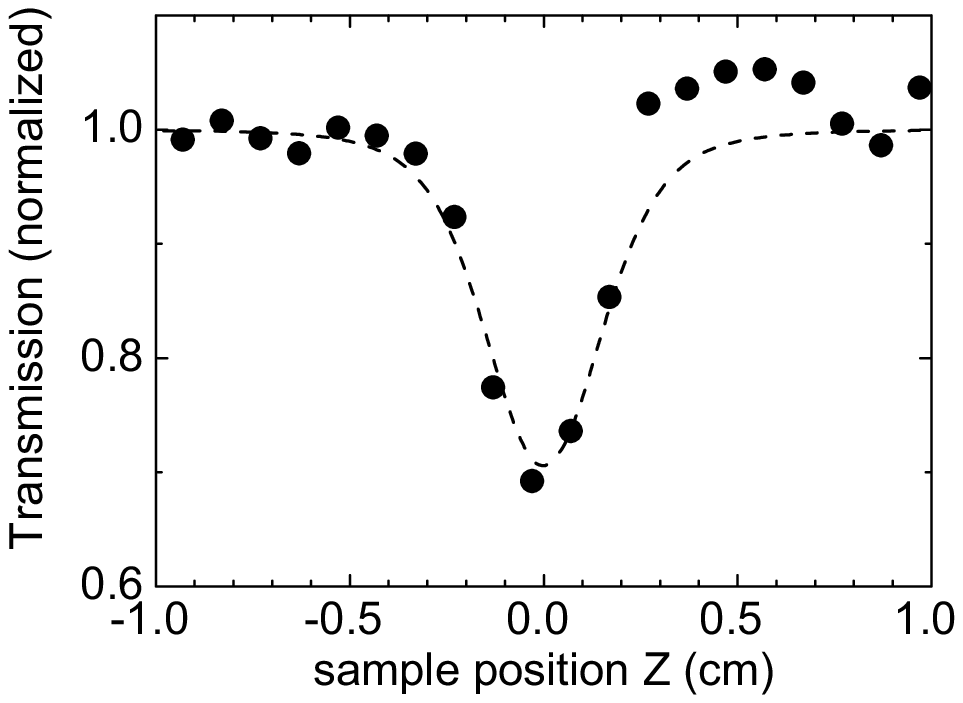}(B). \includegraphics[width=0.4\linewidth]{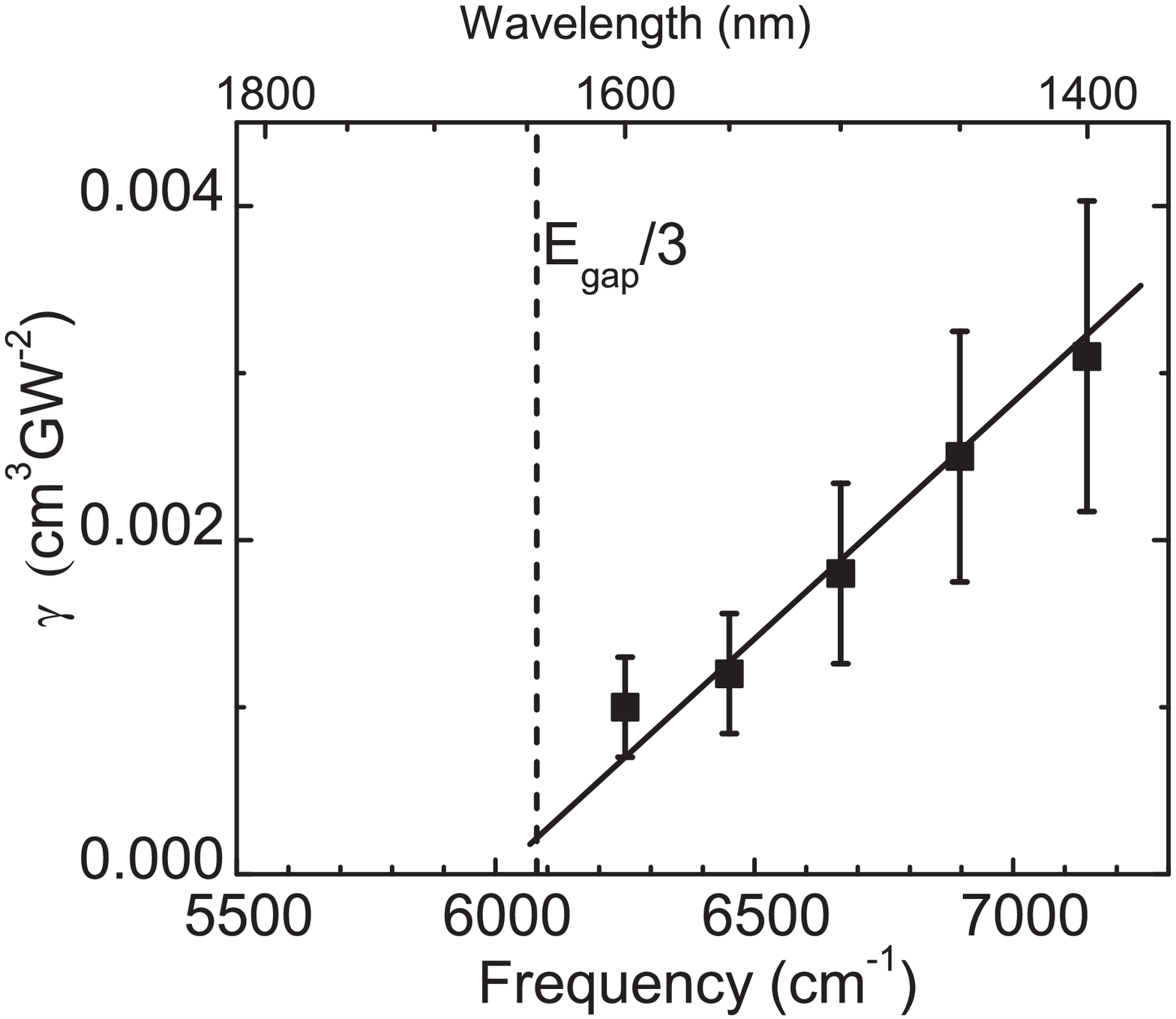}\\
\caption{\label{fig:zscanGaP} (A). Open aperture z-scan measurement for a 300 $\mu$m thick double-side polished GaP
wafer. Pump parameters: $\lambda$= 1600 nm, f= 100 mm, $\tau_p$= 130 fs, $I_0$= 285$\pm$60 GWcm$^{-2}$. The curve
represents the calculated transmission using a three-photon coefficient $\gamma$= 1.0$\pm$0.3$\times$10$^{-3}$. (B). Three-photon coefficient $\gamma$ for GaP were obtained at five wavelengths. The dashed vertical line indicates the three-photon absorption edge for GaP $\frac{1}{3}E_{gap}$, the solid line serves to guide the eye. We observe that $\gamma$ decreases as the pump frequency approaches $\frac{1}{3}E_{gap}$.}
\end{center}
\end{figure}

Fig.~\ref{fig:zscanGaP}(a) shows typical z-scan data taken at a wavelength $\lambda$= 1600 nm, close to the three-photon absorption edge of GaP.  We observe that the z-scan data in Fig.~\ref{fig:zscanGaP}(a) is asymmetric; the transmission of the sample is slightly elevated at positive z-values, where the sample is located in between the beam waist and the detector. This asymmetry of less than 5$\%$ indicates that nonlinear refraction also plays a role in this experiment, and represents a slight deviation from a true open aperture z-scan. From the shape of the curve we conclude that the sign of the nonlinear refraction in GaP is positive in the wavelength range 1400-1600 nm. We therefore exclude that the asymmetry is caused by free-carriers generated by three-photon absorption, since this would result in a negative refractive index change.

To obtain the three-photon absorption coefficient $\gamma$ for GaP, we disregard the relatively small nonlinear refraction. We compare our results to a numerically calculated transmission curve, shown in Fig.~\ref{fig:zscanGaP}(a).  We have varied $\gamma$ until the depth of the minimum of the calculated scan matches the data. The calculated curve agrees well with the data. At $\lambda$= 1600 nm, we deduce $\gamma$= 1.0$\pm$0.3$\times$10$^{-3}$ cm$^3$GW$^{-2}$. Four additional scans were made at $\lambda$= 1400 nm, $\lambda$= 1450 nm, $\lambda$= 1500 nm, and at $\lambda$= 1550 nm.  The resulting deduced three-photon absorption coefficients are plotted versus frequency in Fig.~\ref{fig:zscanGaP}(b). We observe that $\gamma$ tends to zero as the frequency approaches the $\frac{1}{3}E_{gap}$, similar to what was observed in Ref.~\onlinecite{Pearl08} for Si. The frequency scaling confirms a three-photon absorption process. We propose that the observed opposite sign of the negative refractive index change by free-carrier effects and the positive change due to nonlinear refraction in GaP allows for intricate non-monotonic temporal switching of GaP photonic crystals.\cite{Tjerkstra08} This effect would allow ultrafast back-and-forth switching of photonic gaps.\cite{Harding08}

The results of the z-scan measurements that were performed on Si and GaAs in the frequency range close to half the electronic bandgap are summarized in Table~\ref{table:TPA}. The Si data are in good agreement with Refs.~\onlinecite{Dinu03,Bristow07}. The GaAs data are in excellent agreement with Ref.~\onlinecite{Hurlbut07}. The experiments are described in detail in Appendix~\ref{appendix:zscan}.

\begin{table}[h]
 \caption{\label{table:TPA}Two-photon absorption coefficients for Si and GaAs}
 \begin{tabular*}{0.375\linewidth}{lrl}
 \hline \hline
 Material &~~~ $\lambda$ [nm] & $~~~\beta$ [cmGW$^{-1}$]\\
 \hline
 Si   & 1630 & ~~~0.6$\pm$ 0.3\\
 Si   & 1720 & ~~~0.2$\pm$0.1\\
 Si   & 2000 & ~~~0.2$\pm$0.05\\
 GaAs & 1630 & ~~~3.5$\pm$1.0\\
 GaAs & 1720 & ~~~1.5$\pm$0.5\\
 \hline
 \end{tabular*}
 \end{table}

\section{Conclusions}
We have built a two-color pump-probe setup that provides high energy, ultrashort laser pulses at optical frequencies in the range between 3850 and 21050 cm$^{-1}$. Our versatile measurement scheme automatically subtracts the pump background from the probe signal and compensates for pulse-to-pulse variations in the output of our laser. We deduce a tenfold improvement of the precision of the setup, allowing a measurement accuracy of better than $\Delta$R= 0.07$\%$ in a 1 s measurement time, even if the laser is not running optimally. Demonstrations of the technique are presented for a bulk Si wafer, 3D Si inverse opal photonic bandgap crystal, GaAs/AlAs photonic structures, and z-scan measurements on bulk semiconductors.

\section*{ACKNOWLEDGMENTS}
We thank Cock Harteveld, Frans Segerink, and Rindert Nauta for technical support, Soile Suomalainen and Mircea Guina for the Bragg stack sample, the group of David Norris for the Si inverse opal, and Mischa Bonn for useful remarks. This work is part of the research program of the "Stichting voor Fundamenteel Onderzoek der Materie" (FOM), which is supported by the "Nederlandse Organisatie voor Wetenschappelijk Onderzoek" (NWO). WLV thanks FOM for a "Inrichting leerstoelpositie" grant, and support from NWO/VICI and STW/NanoNed.

\section*{Appendix}
\label{appendix:zscan}

\subsection{Two-photon absorption in Si}

\begin{figure}[!ht]
\begin{center}
\includegraphics[width=0.7\linewidth]{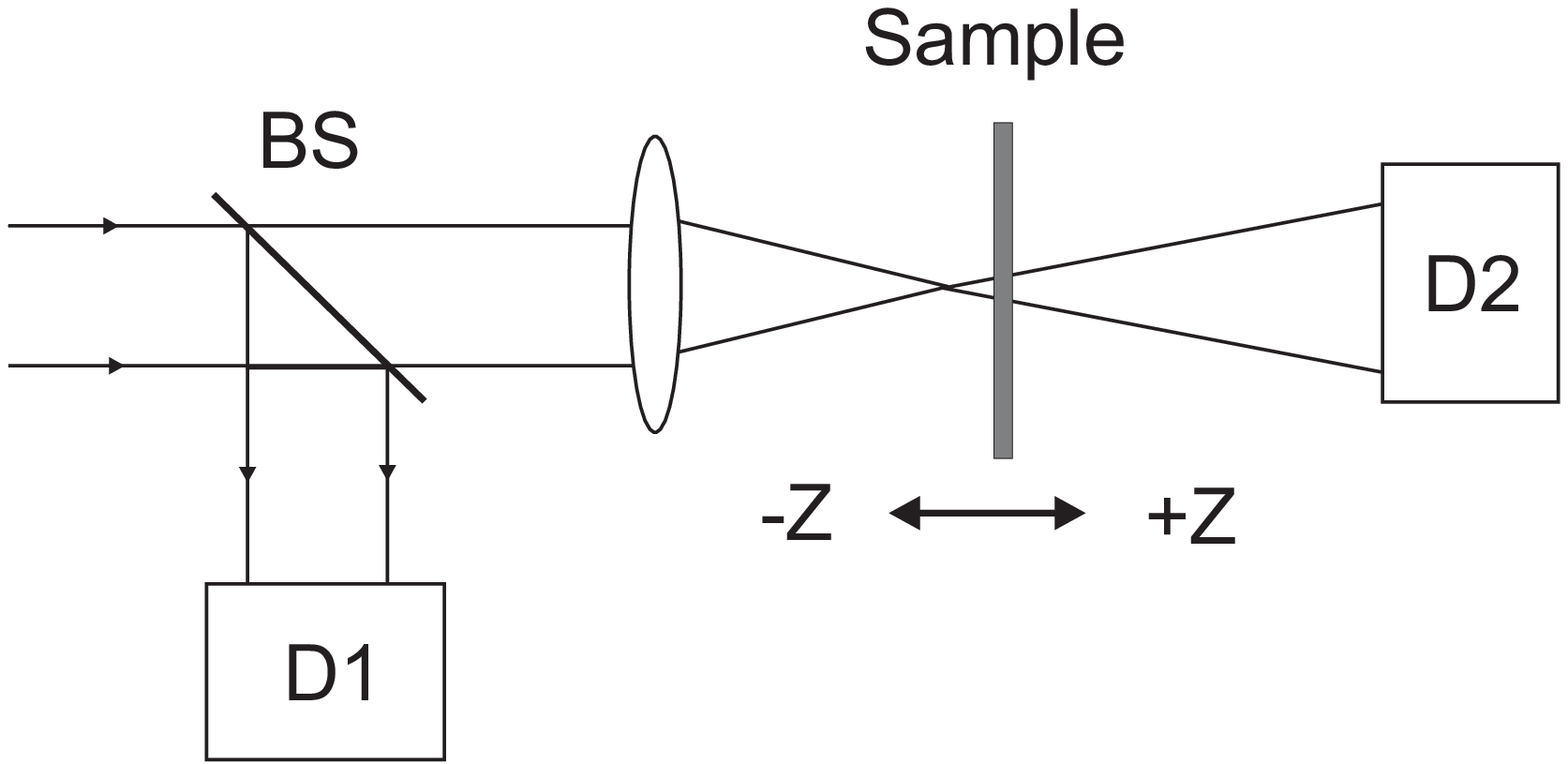}\\
\caption{\label{fig:zscansetup}(Color online) Schematic drawing of the z-scan setup. Incoming laser beam from our OPA is split by
a beam splitter (BS). Two InGaAs photodiodes are used to monitor the output variation of the OPA (D1) as well as the
transmitted signal (D2). The detector signals are measured as a function of sample position $z$.}
\end{center}
\end{figure}

An elegant method to measure the nonlinear refraction $n_2$ of semiconductors is the z-scan technique that was first demonstrated by Sheik-Bahae \emph{et al.}.\cite{Sheikbahae89,Sheikbahae90b} A z-scan is a simple and robust measurement of the transmission of a single beam that is focused by a lens. The focal length of the lens is chosen such that the focal depth is much larger than the sample thickness. The transmitted power of a focused laser beam is measured while the sample is scanned in the z-direction, along the optical axes of the beam, see Fig.~\ref{fig:zscansetup}. The nonlinear refraction results in both Kerr-lensing due to the change in real part of the refractive index, as well as attenuation of the transmitted power due to nonlinear absorption. Both effects will attain a maximum if the sample is located in the beam waist (z=0).

To separate the refractive and absorptive effects, two scans must be made. The first experiment is a \emph{closed aperture} z-scan, in which refraction and absorption are measured simultaneously. A diaphragm that is placed in front of the detector blocks part of the linearly transmitted light. For example, a positive $n_2$ will induce a positive Kerr lens in the sample, which will guide more light into the detectors if the sample is placed at a positive z-position in between the beam waist and detector (see Fig.~\ref{fig:zscansetup}). The transmitted intensity will be reduced if the sample is placed in between the lens band beam waist. The nonlinear refraction will cause an asymmetry in the z-scan data. The second experiment is a an \emph{open aperture} z-scan, in which the aperture is removed, and all transmitted light is collected. The effect of nonlinear refraction is thus removed. The resulting curve only depends on the induced absorption, and is therefore symmetric around the beam waist (z=0). By subtracting the open aperture data from the closed aperture data, the refractive and absorptive effects can be separated.\cite{Sheikbahae90a,Sheikbahae91,Dinu03,Balu04}

In our switching experiments we are mostly interested in the nonlinear absorption coefficients near the two-photon absorption edge of Si and GaAs. Therefore this Appendix describes open aperture z-scan experiments that were done on Si and GaAs single crystalline wafers. To the best of our knowledge the results presented here are the first measurements of the two-photon absorption coefficient GaAs near the two photon absorption edge ($\hbar\omega\approx\frac{1}{2}E_{gap}$).

Fig.~\ref{fig:zscansetup} shows a schematic drawing of our z-scan setup.  The power of the incoming beam is monitored
by detector D1. The beam is focused by a lens with focal length $f$. The sample, typically a double-side polished
wafer, is placed on a translation stage that scans the sample along the z-direction. The zero position is taken at
the beam waist. The transmitted power is collected by a second lens (not shown) and measured by a InGaAs photodiode
D2.

The shape of the transmission curve strongly depends on pump irradiance, and pump power stability is therefore
essential for the correct interpretation of the experimental data. Each pulse was therefore measured
individually using the detection scheme described in Section~\ref{setup:optical}. Both pump-monitor detector
signals (D1) and transmission signals (D2) for each pulse event are stored. We minimize the effect of pulse-to-pulse
variations in the laser output (which can be as high as 10$\%$) by selectively removing all pulses with energy
beyond a certain threshold. In all experiments, the number of pulses collected and the threshold were chosen such that
the standard deviation in pump energy remained below 3$\%$. Typically between 2500 and 10000 pulses were collected for
each datapoint. The corresponding standard error in the transmission was typically better than $\delta$T/T$<$1$\%$,
which is sufficiently sensitive for our z-scan measurements. The sensitivity can be further increased by
narrowing the pulse energy range, at the price of longer integration times.

\subsection{Model}

To interpret the z-scan data, we have numerically calculated the nonlinear transmission of a Gaussian beam
through a thin slab. The beam radius of a diffraction limited Gaussian beam is

\begin{equation}
\label{eq:beamwaist} w(z)=w_0\sqrt{{\Big[}1+{\Big(}\frac{\lambda
z}{\pi w_0}{\Big )}^2{\Big]}},
\end{equation}

\noindent where $\lambda$ is the wavelength, z is the sample position relative to the focus.\cite{Demtroder02} The diffraction
limited radius of the beam waist is equal to:

\begin{equation}
w_0=\frac{f\lambda}{\pi w_b},
\end{equation}

\noindent  where $f$ is the focal length of the lens and $w_b$ is the Gaussian radius of the unfocused beam at the
position of the lens. At each wavelength, $w_b$ was determined by a knife edge scan.

In our calculation we have discretized the sample into 256$\times$256 independent transmission channels of
10$\times$10 $\mu$m$^2$. We have made sure that the lateral dimensions of each channel are smaller than the focus radius
$w_0$, to avoid discretization artifacts at z=0. We calculate the nonlinear
transmission through each channel. Since our pump frequencies are in the two-photon absorption range
$\frac{1}{2}E_{gap} < \hbar\omega < E_{gap}$ where $\alpha$=0, therefore, the transmitted power through a sample with thickness L, normalized to the linear transmission $(1-R)^2$ is equal
to:

\begin{equation}\label{eq:twophoton}
T(z)= \frac{1}{1+ I_0(z) \beta L},
\end{equation}

\noindent where R is the Fresnel reflectance at the front- and back-face of the sample, $\beta$ is the two-photon
absorption coefficient, and $I_0(z)$ the irradiance at the sample interface after subtraction of the front-face
reflection at position z. The added calculated transmission of all channels is plotted versus sample position z. The
adjustable parameter in this calculation is $\beta$.

\begin{figure}[!ht]
\begin{center}
\includegraphics[width=0.7\linewidth]{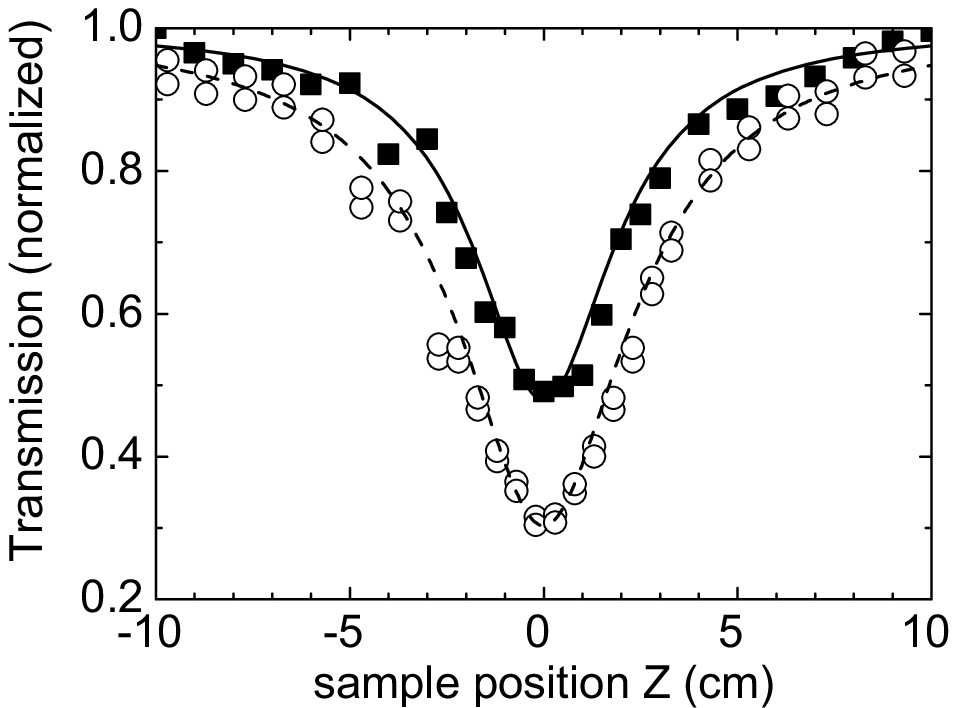}\\
\caption{\label{fig:zscanSi1} Open aperture z-scan measurement for a 360 $\mu$m thick double-side polished Si
wafer. Open circles: $\lambda$= 1630 nm, f= 100 mm,  $I_0$= 385$\pm$40 GWcm$^{-2}$. Closed squares: circles, $\lambda$=
1720 nm, f= 100 mm, $I_0$= 315$\pm$40 GWcm$^{-2}$. The curves are calculated transmission using $\beta$=
0.6$\pm$0.3 cmGW$^{-2}$ (dashed curve, $\lambda$= 1630 nm) and $\beta$= 0.2$\pm$0.1 cmGW$^{-2}$ (solid curve,
$\lambda$= 1720 nm)}
\end{center}
\end{figure}

\begin{figure}[!ht]
\begin{center}
\includegraphics[width=0.7\linewidth]{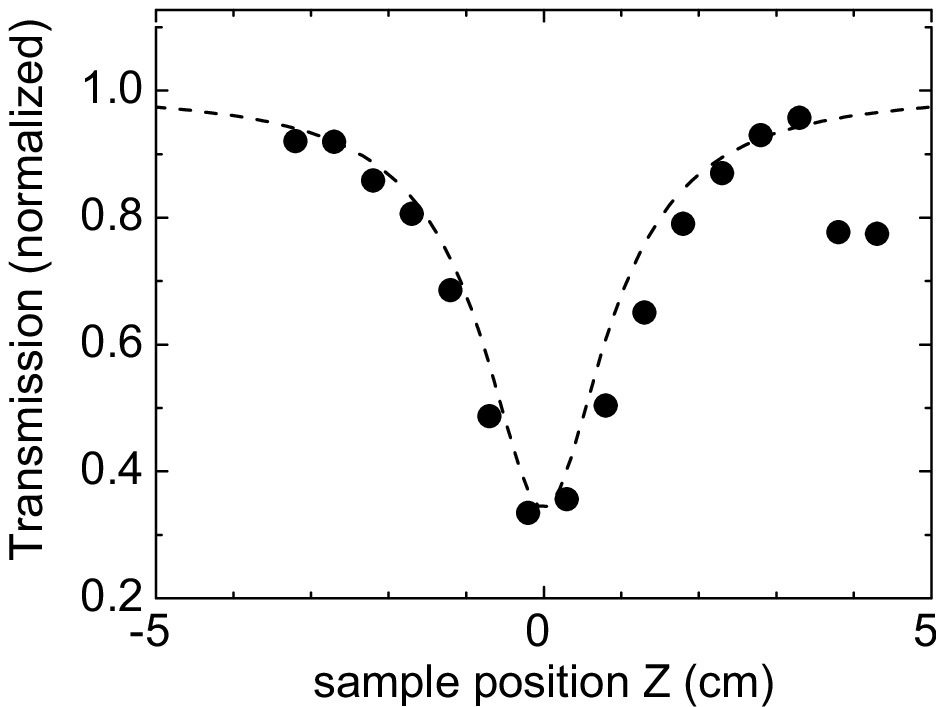}\\
\caption{\label{fig:zscanSi2} Open aperture z-scan measurement for a 360 $\mu$m thick double-side polished Si
wafer. Pump parameters: $\lambda$= 2000 nm, f= 150 mm, $\tau_p$= 130 fs, $I_0$= 800$\pm$200 GWcm$^{-2}$.  The
dashed curve represent the calculated transmission using $\beta$= 0.20$\pm$0.05 cmGW$^{-1}$}
\end{center}
\end{figure}

First we consider open aperture z-scan measurements on a double-side polished Si wafer with thickness L= 360 $\mu$m.
Normalized transmission is plotted in Fig.~\ref{fig:zscanSi1} as a function of sample position z. Data were taken at two wavelengths: $\lambda$= 1630 nm (open circles) and at $\lambda$= 1720 nm (closed squares). The measured data were normalized to the linear transmission away from the focus. We observe that in both scans, the transmission is strongly reduced as the sample scans
through the focus. The curves are numerically calculated transmission data, where $\beta$ was used as fitting
parameter. We find good agreement for $\beta$= 0.6$\pm$0.3 cmGW$^{-2}$ ($\lambda$= 1630 nm), and for $\beta$=
0.2$\pm$0.1 cmGW$^{-1}$ at $\lambda$= 1720 nm. Fig.~\ref{fig:zscanSi2} shows a z-scan of the same Si wafer at a
pump wavelength $\lambda$= 2000 nm, close to the two-photon absorption edge of Si. The peak irradiance during this scan
was $I_0$= 800$\pm$200 GWcm$^{-2}$.  We find good agreement for $\beta$= 0.20$\pm$0.05 cmGW$^{-1}$.

\subsection{Two-photon absorption in GaAs}

\begin{figure}[!ht]
\begin{center}
\includegraphics[width=0.7\linewidth]{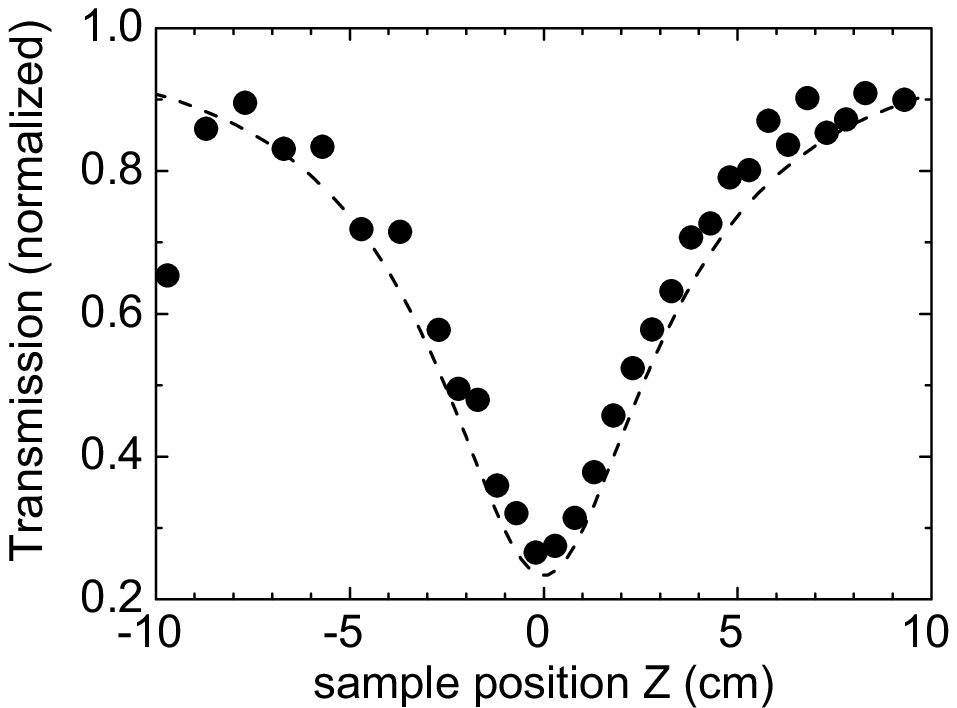}\\
\caption{\label{fig:zscanGaAs17201} Open aperture z-scan measurement for a 189 $\mu$m thick double-side polished
GaAs wafer. Pump parameters: $\lambda$= 1720 nm, f= 100 mm, $\tau_p$= 130 fs, $I_0$= 366$\pm$60 GWcm$^{-2}$.  The curve
represents the calculated transmission using $\beta$= 1.5$\pm$0.5 cmGW$^{-1}$}
\end{center}
\end{figure}

\begin{figure}[!ht]
\begin{center}
\includegraphics[width=0.7\linewidth]{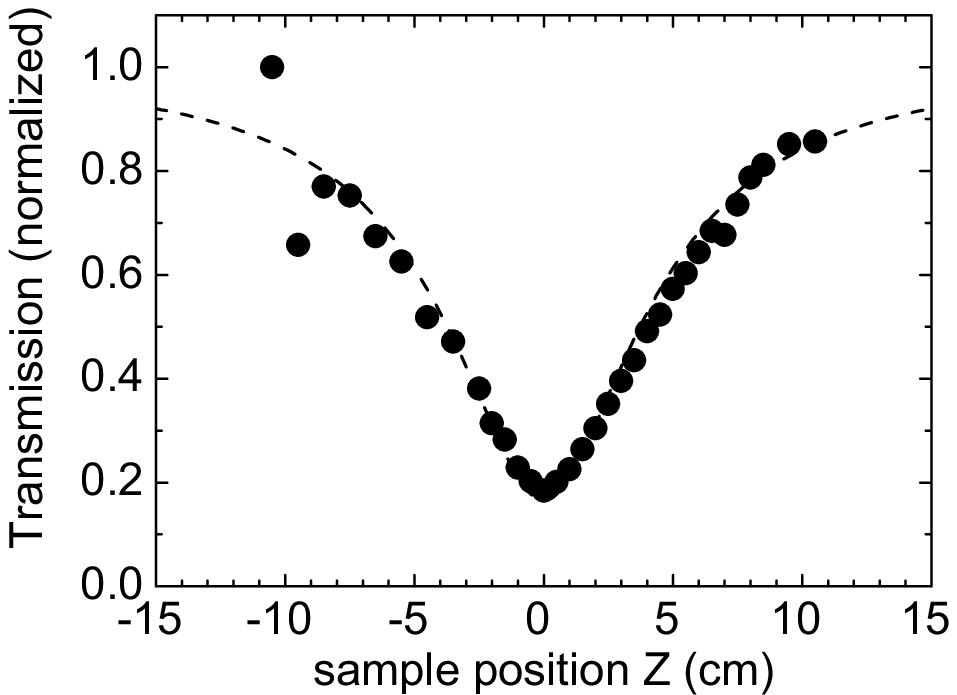}\\
\caption{\label{fig:zscanGaAs17203} z-scan measurement for a 189 $\mu$m thick double-side polished GaAs wafer. Pump
parameters: $\lambda$= 1630 nm, f= 100 mm, $\tau_p$= 130 fs, $I_0$= 244$\pm$40 GWcm$^{-2}$.  The curve represents
the calculated transmission using $\beta$= 3.5$\pm$1.0 cmGW$^{-1}$}
\end{center}
\end{figure}

We have performed open aperture z-scan experiments on a double-side polished GaAs wafer with a
thickness of 189 $\mu$m. Figure~\ref{fig:zscanGaAs17201} shows z-scan data taken at $\lambda$= 1720 nm, just above
the two-photon absorption edge of GaAs. The measured data were normalized to the linear transmission away from the
focus. The peak irradiance was $I_0$= 366$\pm$60 GWcm$^{-2}$ in Fig.~\ref{fig:zscanGaAs17201}. The strongly
attenuated transmission near the waist of the beam indicates a strong nonlinear absorption. The curve is
calculated using $\beta$ as an adjustable parameter. We find good agreement for $\beta$= 1.5$\pm$0.5 cmGW$^{-1}$.
Fig.~\ref{fig:zscanGaAs17203} shows z-scan data for GaAs at a shorter wavelength $\lambda$= 1630 nm. Here, we find good
correspondance for $\beta$= 3.5$\pm$1.0 cmGW$^{-1}$. Our data are in excellent agreement with Ref.~\onlinecite{Hurlbut07}. We conclude that in GaAs, the two-photon absorption coefficient strongly decreases as the pump wavelength approaches half the band gap energy, allowing spatially more homogeneous switching.\cite{Euser05}

We have measured the two-photon absorption coefficients of Si and GaAs near the two-photon absorption edge by an open-aperture z-scan technique. The experimental data was compared to a model that includes nonlinear absorption in the sample. For both Si and GaAs we find that the two-photon absorption coefficient tends to zero near half the gap energy $\frac{1}{2}E_{gap}$.
\end{document}